\documentclass[page-classic]{epl2}
\usepackage{amssymb}

\title{Graviton resonances on deformed branes}
\shorttitle{Graviton resonances on deformed branes} 

\author{W. T. Cruz\inst{1} \and A. R. Gomes\inst{2} \and C. A. S. Almeida\inst{3}}

\shortauthor{W.T. Cruz \etal}

\institute{
\inst{1}Instituto Federal de Educa\c{c}\~{a}o, Ci\^{e}ncia e Tecnologia do Cear\'{a} (IFCE), Campus Juazeiro do Norte - 63040-000 Juazeiro do Norte-Cear\'{a}-Brazil\\
\inst{2} Instituto Federal do Maranh\~ao, Campus Monte Castelo, S\~ao Lu\'is - Maranh\~ao - Brazil \\
\inst{3} Departamento de F\'{i}sica - Universidade Federal do Cear\'{a} - C.P. 6030, 60455-760 Fortaleza - Cear\'{a}-Brazil}


\abstract{ Plane wave solutions of Schrodinger-like equations obtained from the metric perturbations in 5D braneworld scenarios can present resonant modes. The search for those structures is important  because they can provide us massive modes with not suppressed couplings with the membrane. We propose in this paper the study of graviton Kaluza-Klein spectrum in a special kind of membrane that possesses internal structure. The interest in study of these deformed defects is because they have a more rich internal structure that has implications in the matter-energy density along the extra dimensions an this produces a space-time background whose curvature has a splitting, if compared to the usual kink-like models. Such models arise from $(4,1)$-branes constructed with one scalar field coupled with gravity where we find two-kink solutions from deformations of a $\phi^4$ potential. The main objective of this work is to observe the effects of deformation process in the resonant modes as well as in the coupling between the graviton massive modes and the brane.}

\begin{document}

\maketitle

\section{Introduction}

The study of topological defects in braneworld scenarios has increased recently due to its advantages over the Randall-Sundrum model \cite{rs1,rs2} in warped geometries. In extra dimension scenarios we can represent the observable universe as a hypersurface embedded in a multidimensional space. Having appeared as a proposal to solution of the gauge hierarchy problem, in such models particles of the Standard Model must be confined in the four dimensional brane.

In the seminal works of Bazeia and collaborators \cite{defects1,defects2,defects3, aplications} a class of defect structures were obtained by a $\phi^4$ potential. Such structures were called 2-kink defects since they seem to be composed of two standard kinks, symmetrically separated by a distance which is proportional to the deformation parameter p. When applied to warped geometries with one extra dimension such topological deformed defects are used to mimic braneworlds where the observable universe are located. The scalar field is the stuff the brane is made of. In deformed thick brane models, as we have used in this work, the region between the two interfaces $(\phi=\pm1)$ of the defect are more richer than the usual kink solution. It is important to note that despite the emergence of a new structure between the two minimum of the potential at $\phi=\pm1$ and that this new structure seems to be two separate kinks, we are still working with a single deformed structure. The same feature was used in braneworld models to describe the splitting of a thick domain wall \cite{fase}. To braneworld models the appearance of a new region inside the bounce solution due to the deformation procedure will result in the appearance of a gap in the matter energy density in the center of the brane, suggesting the appearance of internal structure \cite{bloch, adalto}. In models with  five-dimensional gravity coupled to scalars, we can obtain bounce solutions with smooth metric warp functions \cite{de,csaba,gremm,fase,bc,kehagias}. Such thick brane solutions are more natural because they are dynamically generated by a $\phi^4$ potential.


The scenario that we use here was initially described in \cite{adalto,deformed} where a class of topological defect solutions was constructed starting from a specific deformation of the $\phi^4$ potential. These new solutions
may be used to mimic new brane-worlds containing internal structures. Such internal structures have implications in the density of matter-energy along the extra dimensions \cite{bloch} and this produces a space-time background whose curvature has a
splitting. Some characteristics of such model were considered in phase transitions
in warped geometries \cite{fase}. A series of discussions about splitting branes and its applications to condensed matter are found in \cite{cond}.

We consider the study of massive modes and resonances of graviton in branes generated by deformed defects. The center of the defect, where the observable universe is localized, is considered in $y=0$ where $y$ represents the extra dimension. The same feature was considered in our previous works in order to analyze resonances for gauge fields \cite{wilami_e_ca1}, Kalb-Ramond field \cite{wilami_e_ca2}, and fermions fields\cite{ca} in different geometries. Specifically in the Ref. \cite{ca} some of the authors of the present work were the first to consider fermionic resonances in branes with internal structure. We can also cite the works \cite{Liu4, Liu5} that consider the study of resonances of fermionic fields. These reviews are very interesting since they present branes with splitting which is observed by the division in the minimum of the Shroedinger potential, where a series of resonance structures are found. Similar characteristics are encountered in the present work.

As noticed in for some values of mass, the plane wave solutions of Schrödinger-like equations obtained in the transverse-traceless (TT) sector of metric perturbations can present very high amplitudes inside the brane.  We can interpret this as resonant modes and the existence of these structures can give us a KK spectra with not suppressed coupling with the matter inside the membrane. Graviton resonances were previously considered in RS background \cite{csaba1} being related with the existence of scales on which the gravitational laws appear to be four dimensional. In this scenario the width of the graviton resonance gives the lifetime. More recently, several papers have detected resonant modes in the study of localization of gauge field \cite{wilami_e_ca1}, Kalb-Ramond field \cite{wilami_e_ca2} and fermionic fields \cite{Liu4,ca} in branes with internal structure.

The main objective of this work is to study the behavior of gravity in a membrane generated by a deformation procedure. The so-called two-kink solutions, that are the stuff the branes is made of, can be obtained after a deformation procedure of a potential from a scalar field \cite{deformed}. Using the resonance detecting method described in \cite{Liu4,ca,wilami1,wilami_e_ca1,wilami_e_ca2}, we analyze the massive modes arising from the dimensional reduction functions. As we will see, the internal structures given by the deformations on the brane will have implications to the coupling of massive modes to the brane. Such influences will affect the resonance structures found.

This paper is organized as follows: in the second section we describe the D = 5 space-time background and describe the deformed membrane setup. In the third section we study wave solutions of Schrodinger-like equations given by metric perturbations. The following section is devoted to our conclusions and perspectives.

\section{Brane setup}

We start with the action describing one scalar field minimally coupled with gravity in five dimensions
\begin{equation}
S=\int d^{5}x \sqrt{-G}[2M^{3}R-\frac{1}{2}(\partial\phi)^{2}-V(\phi)],
\end{equation}
where $\phi$ is the scalar field, the stuff making the membrane, $M$ is the Mass Planck in $D=5$ dimensions and $R$ is the  scalar curvature.

For some classes of the potential $V(\phi)$, it is possible to obtain kink solutions for the field $\phi$ depending only on the extra dimension.

As an ansatz for the metric we consider an extension for the Randall-Sundrum metric, where the bulk spacetime is asymptotically $AdS_5$, with a Minkowski brane,
\begin{equation}\label{metrica}
ds^{2}=e^{2A(y)}\eta_{\mu\nu}dx^{\mu}dx^{\nu}+dy^{2}.
\end{equation}
The scalar field and the warp factor depend only on the extra dimension $y$.
The tensor $\eta_{\mu\nu}$ is the Minkowski metric and the indices
$\mu$ and $\nu$ vary from 0 to 3. For this background we find the following equations of motion:
\begin{equation}
\phi^{\prime\prime}+4A^\prime\phi^\prime=\frac{dV}{d\phi},
\end{equation}
\begin{equation}\label{mov1}
\frac{1}{2}(\phi^{\prime})^{2}-V(\phi)=24M^{3}(A^{\prime})^{2}.
\end{equation}
and
\begin{equation}\label{mov2}
\frac{1}{2}(\phi^{\prime})^{2}+V(\phi)=-12M^{3}A^{\prime\prime}-24M^{3}(A^{%
\prime})^{2}.
\end{equation}
Here prime means derivative with respect to the extra dimension.

In the presence of gravity, defining the potential as
\begin{equation}
V_p(\phi)=\frac{1}{2}\left(\frac{dW}{d\phi}\right)^2-\frac{8M^3}{3}W^2,
\end{equation}
it is possible to find first-order equations
\begin{equation}
\label{1order}
\phi'=\frac{\partial W}{\partial \phi},
\end{equation}
\begin{equation}
\label{WA}
W=-3A'(y),
\end{equation}
whose solutions are also solutions from the equations of motion. Here $W(\phi)$ is the superpotential. This formalism was initially introduced in the study of non-supersymmetric domain walls in various dimensions \cite{de,sken}.
For bounce-like solutions, the field $\phi$ tends to different values when $y\rightarrow\pm\infty$. Such solutions can be attained by a double-well potential. In this way, guided by refs. \cite{new,defects-inside,deformed,adalto}, we chosen the superpotential,
\begin{equation}\label{sup}
W_p(\phi)=\frac{p}{2p-1}\phi^{\frac{2p-1}{p}}-\frac{p}{2p+1}\phi^{\frac{2p+1}{p}},
\end{equation}
where the parameter $p$ is an odd integer. The chosen form for $W_p$ was constructed after deforming the $\lambda\phi^4$ model. This choice allows us to obtain well-defined models when $p=1,3,5,...$, where for $p=1$ we get the standard $\phi^4$ potential. For $p=3,5,7,...,$ the potential $V_p$ presents a minimum at $\phi=0$ and two more minima at $\pm 1$. Eq. (\ref{1order}) can be easily solved giving the so called two-kink solutions
\begin{equation}\label{twokink}
\phi_p(y)=\tanh^p\biggl(\frac{y}{p}\biggr).
\end{equation}
%
%
%
%
%
we can find  explicitly the solution for $A_p(y)$  as \cite{adalto},
%
%
\begin{eqnarray}\label{a}
A_p(y)=-\frac{1}{3}\frac{p}{2p+1}\tanh^{2p}\left(\frac{y}{p}\right)- \frac{2}{3}\left(\frac{p^2}{2p-1}-\frac{p^2}{2p+1}\right)\\\nonumber
\biggl{\{}\ln\biggl[\cosh\left(\frac{y}{p}\right)\biggr]- \sum_{n=1}^{p-1}\frac1{2n}\tanh^{2n}\left(\frac{y}{p}\right)\biggr{\}},
\end{eqnarray}

\begin{figure}[ht!]
\includegraphics[width=13.5cm,height=4.5cm]{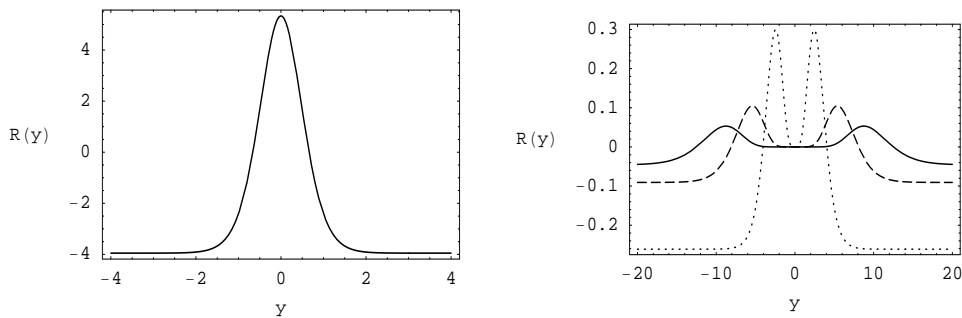}
\caption{\label{curvp}Plots of the solution of the curvature invariant $R(y)$ for $p=1$ on the left. On the right for
$p=3$ (doted line), $p=5$ (dashed line) and $p=7$ (solid line).}
\end{figure}
The parameter $p$ is an odd integer. The chosen form for $W_p$ allows us to obtain well-defined models when $p=1,3,5,...$, where for $p=1$ we get the standard $\phi^4$ potential.
Note that the exponential warp factor constructed with this function is localized around the membrane and for large $y$ it approximates the Randall-Sundrum solution \cite{rs2}.

In addition we present the shape of the scalar curvature, which due to the metric shape does not blows up to infinity, which is another advantage of this model. We can also observe the splitting on the scalar curvature generated by the deformations. For instance, the Ricci scalar is
\begin{equation}
R=-[8A_p''+20(A_p')^{2}].
\end{equation}
Fig. \ref{curvp} shows the Ricci scalar for $p=1,3,5,7$. Note that the Ricci scalar is finite at all points in the bulk. Far from the brane $R$ tends to a negative constant, characterizing the $AdS_5$ limit for the bulk. Note that the higher is the parameter $p$, the lower is the constant, interpreted as the inverse of the $AdS$ scale. For $p=1$ the Fig. \ref{curvp} shows that the scalar curvature for a non-deformed model presents a maximum at the brane center $y=0$, whereas for higher values of $p$ this maximum is splitted into two smaller peaks that decrease with the increasing of $p$. The presence of regions with positive Ricci scalar can in principle be connected to the capability to trap massive states near to the brane, as we will investigate in the following sections. Also note that for even larger values of $p$ we see that the tendency is the Ricci scalar to approach to zero near to the brane center. In this way we expect the presence of gravity resonances to be more pronounced for lower values of $p$.

\section{Gravity localization and Resonances}

The issue of gravity localization in this class of models was considered by one of us in Ref. \cite{adalto}. As already noted in \cite{gremm}, plane wave solutions of Schrodinger-like equations in the transverse-traceless sector of metric perturbations can present solutions as resonant modes. Such structures were obtained by Csaki \textit{et al.} \cite{csaba,csaba2} when studying gravity localization.

Now we investigate numerically the presence of resonances with a more refined method. The stability analysis is performed after perturbing
the metric as follows
\begin{equation}
ds^2=e^{2A(y)}(\eta_{\mu\nu}+h_{\mu\nu})dx^\mu dx^\nu-dy^2.
\end{equation}
Here $h_{\mu\nu}=h_{\mu\nu}(x,y)$ are small perturbations. In the transverse-traceless gauge the perturbations turn to ${\bar h}_{\mu\nu}$, and the metric and scalar field fluctuations decouple, resulting in the equation
\begin{equation}
\label{h}
{\bar h}_{\mu\nu}^{\prime\prime}+4\,A^{\prime}
\,{\bar h}_{\mu\nu}^{\prime}=e^{-2A}\,\Box\,{\bar h}_{\mu\nu}.
\end{equation}
Here $\Box$ stands for the 4-dimensional D'Alembertian. The extra dimension $y$ is turned into a new coordinate $z$, defined by
\begin{equation}\label{trans1}
\frac{dz}{dy}=e^{-A_p},
\end{equation}
which makes the metric conformally flat. Also, with the help of redefinition
\begin{equation}
{\bar h}_{\mu\nu}(x,z)=e^{ik\cdot x}e^{-\frac{3}{2}A(z)}H_{\mu\nu}(z),
\end{equation}
as shown by Bazeia \textit{et al.} \cite{adalto,bgl}, the equation for the metric fluctuations takes the form of a Schr\"odinger-like equation
\begin{equation}
\label{se}
-\frac{d^2H_{\mu\nu}}{dz^2}+U_p(z)\,H_{\mu\nu}=k^2\,H_{\mu\nu},
\end{equation}
where the potential is given by
\begin{equation}
U_p(z)=\frac32\,A_p^{\prime\prime}(z)+\frac94\,A_p^{\prime2}(z).
\end{equation}

It was already shown \cite{adalto} that the Hamiltonian is positive definite and tachyonic modes are absent and that it is possible to attain an explicit expression for the non-normalized zero-modes, responsible for gravity localization.

\begin{figure}
\centering
\includegraphics[angle=270,width=8cm]{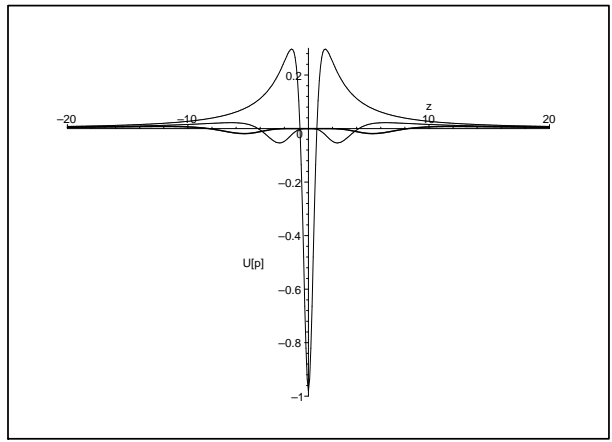}
\caption{Plots of the potentials $U_1(z), U_3(z),\, {\rm and}\, U_5(z)$.}
\label{pot-grav}
\end{figure}

Fig. \ref{pot-grav} shows that the Schrodinger potentials for gravity fluctuations have the form of volcano potentials. This inspired us to investigate the possibility of resonances with the Numerov method \cite{bgl}, identifying resonances as a peak in the normalized squared wavefunctions $|H_{\mu\nu}(0)|^2$ at the brane center.

 We considered normalization in a box with ends at $\pm z_{max}$, far enough for the inverse square law $U_p(z)\sim \alpha_p(\alpha_p+1)/z^2$ to be achieved. As gravity localization can be determined by the far region of the potential \cite{csaba,csaba2}, the plot of $z^2U_p(z)$ for $-200<z<200$ gives $\alpha_1=1.483$. This characterizes gravity localization where $U_{grav}(r)$, the gravitational potential  between two unit masses distant $r$ one from the other, reproduces the Newtonian limit for large distances. However, for short distances there is a $1/r^{2\alpha_p}$ correction due to the small massive modes. For $p=1$ this gives $1/r^{2.966}$, close to the Randall-Sundrum $1/r^3$ correction for the Newtonian potential. For larger values of $p$, one needs larger values of $z_{max}$ to achieve the $1/z^2$ region for $U_p$.
 \begin{figure}
 \centering
\includegraphics[{angle=270,width=6.5cm}]{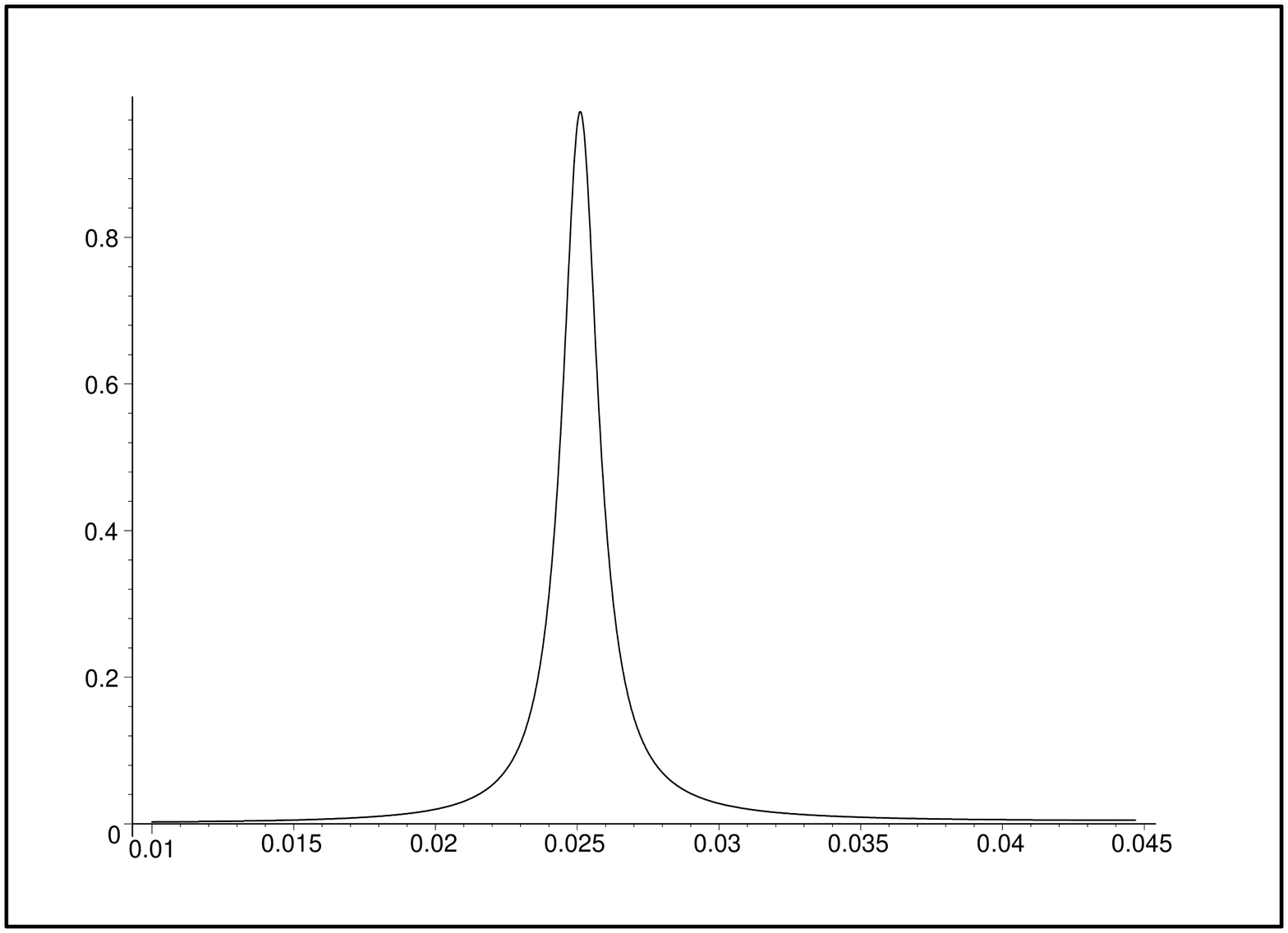}
\includegraphics[{angle=270,width=6.5cm}]{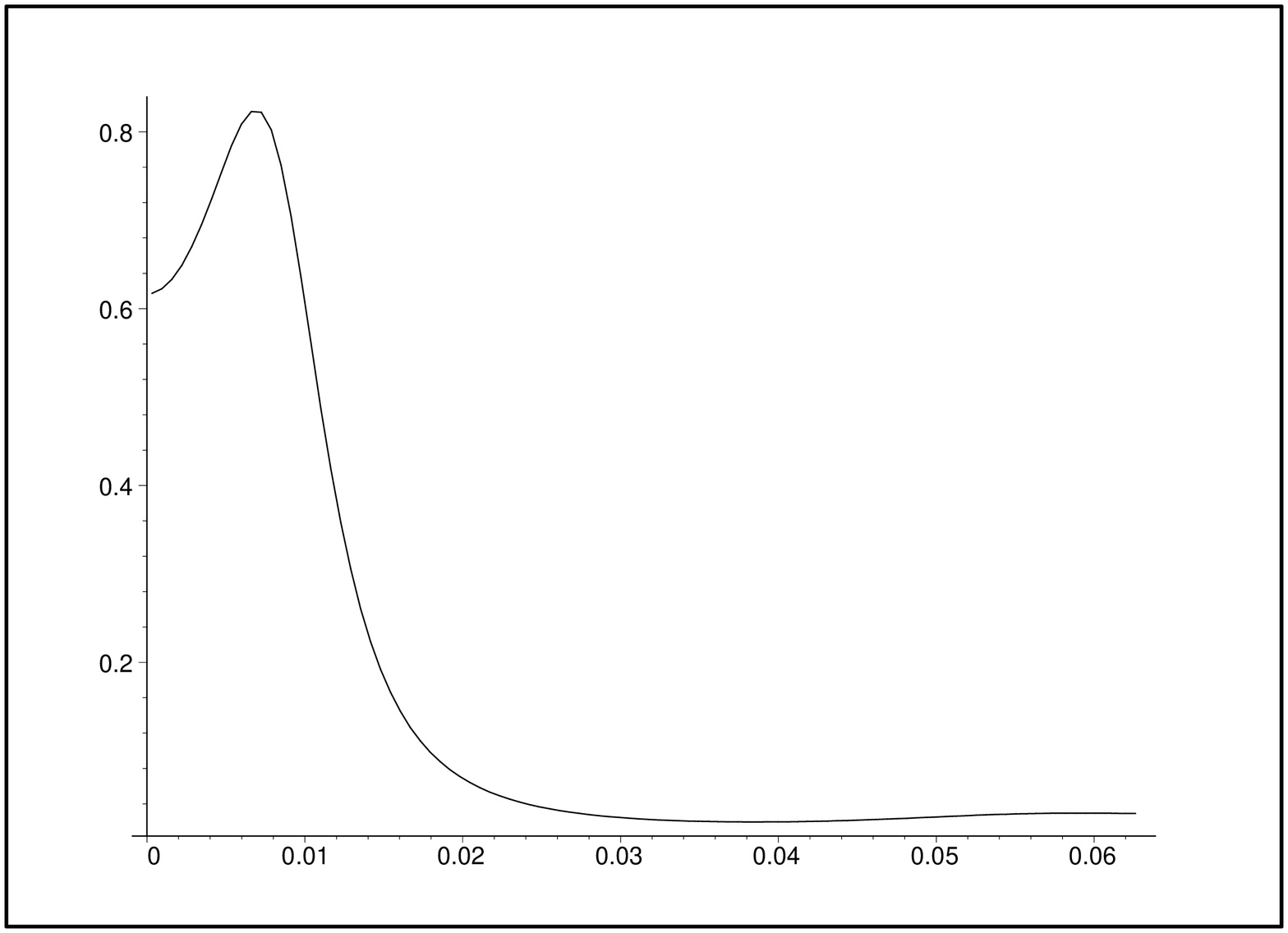}
\includegraphics[{angle=270,width=6.5cm}]{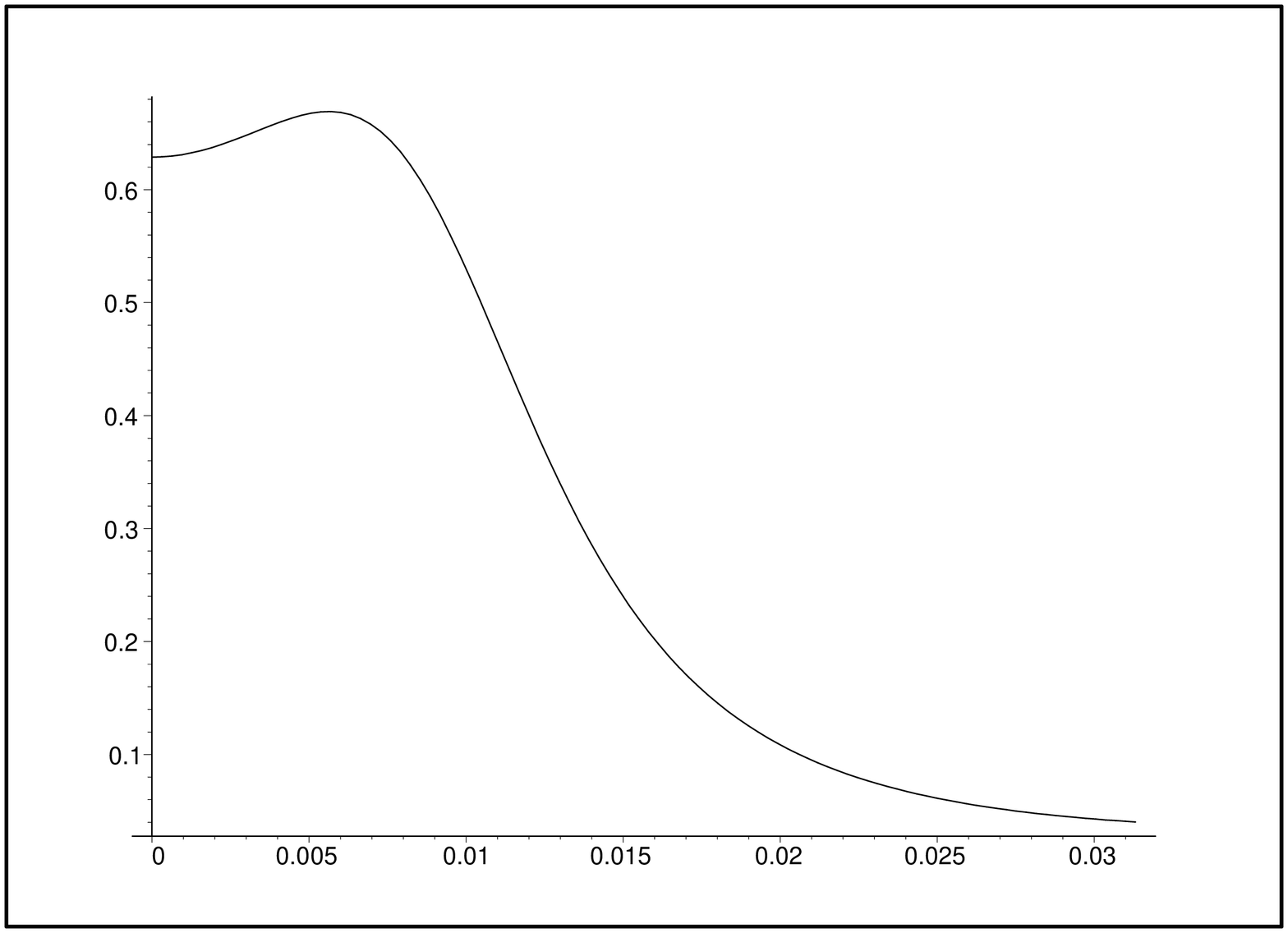}
\caption{Normalized $|H_{\mu\nu}(0)|^2$, as a function of $m$, for $m>0$, showing the resonance peaks for (b) $p=1$ (upper right), $p=3$ (upper left) and $p=5$ (lower).}
\label{H}
\end{figure}

In order to studying resonances, we do not need to persecute large values of $z_{max}$, since we are interested in the effect where the wavefunction behavior changes abruptly for a particular mass. For our purposes it is sufficient to consider a normalization procedure with $z_{max}=100$. We found a clear peak for $p=1$, as showed in Fig. \ref{H}. We noted that with the increasing of $p$, the resonance peaks become broader, showing that branes with smaller values of $p$ are more effective in trapping graviton KK modes. This can also be noted with zero modes, where the modes decay as $1/\sqrt z$ for $z>z_p$ and $z_p$ grow with $p$ (see Ref. \cite{adalto}). For $p=5$ the peak thickness $\Delta m$ is too large for characterizing a resonance. This was expected since the maxima of the Schrodinger potential, firstly pronounced for $p=1$ is reduced considerably for larger values of $p$.

About our numerical method it is worthwhile to mention that we use the method used by one of the authors of the present work in Ref.\cite{bgl}, in order to investigate the spectrum of massive modes and its contribution for gravity localization on thick branes. An important point concerning that method is the exigence of a $Z_2$ symmetric potential and that its applicability is only for even massive modes. Details about that method are reviewed in the Appendix of Ref.\cite{bgl}. On the other hand, recently Cvetic and Robnik \cite{cvetic} showed the presence of an extremely  thin resonances in an integrable model where the graviton wave function modes were explicitly parametrized by the thickness of the wall. The applicability of the numerical method used in the present work could be confronted with the results from Cvetic and Robnik, with excellent quantitative agreement. This agreement was attained even for larger masses, much lower then the value of the potential energy at the top of the "crater of the volcano".

\section{Conclusions}

In this work we have studied the KK modes of gravity in a model of deformed branes. The parameter $p$ of the model controls important features as the brane thickness and energy distribution along the extra dimension. We found that the increasing of $p$ cause a splitting in the Ricci scalar, evidencing the appearance of an internal structure, as can be found also from the energy density analysis (see \cite{adalto}). We investigate metric perturbations as decoupled from the scalar ones in the transverse-traceless gauge. From the asymptotic behavior of the Shr\"odinger-like equation it can be proved that gravity is localized for all parameters $p$, and tachyonic gravity modes are absent. The normalized squared wavefunctions $|H_{\mu\nu}(0)|^2$ that we use to detect modes with high amplitudes in the brane center reveal us that, excluding the resonant modes, in deformed branes the lightest modes couples strongly to the branes in comparison to KK modes with higher masses.

Resonances in the gravity sector are observed for $p=1$ and $3$ as peaks in the $|H_{\mu\nu}(0)|^2$ distribution. However, larger values of $p$ lead to produce broader peaks that cannot technically characterize themselves as resonances. This means that thinner branes are more effective for trapping gravity. The increase of the deformation parameter $p$ gives branes where the splitting is more evident. However when analyzing the difference between the characteristics of the resonances found for $p=1$, $p=3$ and $p=5$ we can observe that when the splitting of the brane increases such structures are destroyed. Similar characteristic was noted in the study of the massive spectrum of vectorial and tensorial gauge fields \cite{wilami_e_ca1,wilami_e_ca2}. One last remark that we make is about the size of the resonances. As claimed in ref \cite{grs} the width of the graviton resonance is inversely connected to the lifetime.  If the resonance is very narrow the lifetime becomes large.  Otherwise if the resonances are broader its effects have no phenomenological relevance.  In our analysis this characteristic appears in brane solutions which exhibit high splitting.

The authors would like to thank FUNCAP, FAPEMA, CNPq and CAPES/PROCAD (Brazilian agencies) for financial support.

\end{document}